\documentclass[journal]{IEEEtran}
\IEEEoverridecommandlockouts
\usepackage{lineno}
\usepackage{cite}
\usepackage{hyperref}
\usepackage{amsmath,amssymb,amsfonts}
\usepackage{amsmath}
\usepackage{amsthm}

\usepackage{graphicx}
\usepackage{textcomp}
\usepackage{xcolor}
\usepackage{graphicx}
\usepackage{float}
\usepackage{subfigure}
\usepackage{amsmath}
\usepackage{amsfonts,amssymb}
\usepackage{mathrsfs}
\usepackage{mathtools}
\usepackage{algorithm}
\usepackage{algorithmicx}
\usepackage{algpseudocode}
\usepackage{bm}
\usepackage{multirow}
\usepackage{array}
\usepackage{amssymb}
\usepackage{amsmath}
\usepackage{cite}
\usepackage{url}
\usepackage{xcolor}
\usepackage{cite,graphicx,amsmath,amssymb}
\usepackage{subfigure}
\usepackage{fancyhdr}
\usepackage{mdwmath}
\usepackage{mdwtab}
\usepackage{caption}
\usepackage{amsthm}
\usepackage{setspace}
\usepackage{bm}
\usepackage{algorithm}
\usepackage{algpseudocode}
\usepackage{mathtools}
\usepackage{dsfont}
\usepackage{bbm}
\newtheorem{remark}{Remark}
\theoremstyle{definition}
\newtheorem{theorem}{Theorem}

\newtheorem{lemma}{Lemma}

\newtheorem{corollary}{Corollary}

\makeatletter
\newcommand{\biggg}{\bBigg@{3}}
\newcommand{\Biggg}{\bBigg@{3.5}}
\makeatother
\begin{document}
\title{On the Performance of Uplink ISAC Systems}
\author{Chongjun~Ouyang, Yuanwei~Liu, and Hongwen~Yang
\thanks{C. Ouyang and H. Yang are with the School of Information and Communication Engineering, Beijing University of Posts and Telecommunications, Beijing, 100876, China (e-mail: \{DragonAim,yanghong\}@bupt.edu.cn).}
\thanks{Y. Liu is with the School of Electronic Engineering and Computer Science, Queen Mary University of London, London, E1 4NS, U.K. (e-mail: yuanwei.liu@qmul.ac.uk). (Corresponding author: Yuanwei Liu)}
}
\maketitle

\begin{abstract}
This letter analyzes the performance of uplink integrated sensing and communications (ISAC) systems where communication users (CUs) and radar targets (RTs) share the same frequency band. A non-orthogonal multiple access (NOMA) protocol is adopted in the communication procedure of the ISAC system. Novel expressions are derived to characterize the outage probability, ergodic communication rate, and sensing rate. Besides, the diversity order and high signal-to-noise ratio (SNR) slope are unveiled to gain further insights. It is found that when achieving the same communication rate, the ISAC system enjoys a higher sensing rate than the conventional frequency-division sensing and communications (FDSAC) system where CUs and RTs share isolated bands. All the results are validated by numerical simulations and are in excellent agreement.
\end{abstract}


\section{Introduction}
Integrated sensing and communications (ISAC) is believed to be a promising spectrum-sharing candidate for future wireless networks \cite{Liu2021}. By integrating the wireless communications and radar sensing to share the same spectrum and infrastructure, ISAC is capable of improving the spectral efficiency, reducing the hardware cost, and limiting the electromagnetic pollution \cite{Liu2021}. These advantages have attracted vibrant industrial and academic interest in the ISAC technique \cite{Liu2021,Zhang2021,Mu2021,Wang2021,Liu2022,Rahman2020,Tang2019}.

Recently, a considerable literature has grown up around the theme of ISAC. For a review, please see the recent papers \cite{Liu2021,Zhang2021,Mu2021,Wang2021,Liu2022,Rahman2020,Tang2019} and references therein. Yet, it is worth noting that most studies in the field of ISAC have only focused on the waveform or beamformig design. In contrast to this, there has been little quantitative analysis of the basic performance of ISAC systems, while only a couple of papers appeared recently \cite{Heath2020,Chiriyath2016}. For example, the communication-sensing performance tradeoffs in ISAC systems with a single communication user (CU) were discussed from an estimation-theoretical perspective by using the Cram\'{e}r-Rao bound metric for radar sensing and minimum mean-square error metric for communications \cite{Heath2020}. On a parallel track, the communication-sensing rate region achieved in a single-CU ISAC system was characterized from an information-theoretical perspective \cite{Chiriyath2016}. In a nutshell, these works have laid a solid foundation for understanding the fundamental performance of ISAC systems. However, in these works, the influence of channel fading as well as multi-user interference was not taken into account and no more in-depth system insights, such as the diversity order and high signal-to-noise ratio (SNR) slope, were unveiled.

To fill this knowledge gap, this letter investigates the performance of uplink multiple-CU ISAC systems where a non-orthogonal multiple access (NOMA) protocol is involved in the communication procedure. We use the outage probability (OP) as well as the ergodic communication rate (ECR) metrics for communications and maximal sensing rate metric for radar sensing. Exact expressions for these metrics as well as their high-SNR approximations are provided. Numerical results suggest that when achieving the same communication rate, the ISAC system yields a higher sensing rate than the frequency-division sensing and communications (FDSAC) system where isolated bands are used for communications and sensing, respectively.

\section{System Model}
\begin{figure}[!t]
    \centering
    \subfigbottomskip=0pt
	\subfigcapskip=0pt
\setlength{\abovecaptionskip}{0pt}
   \subfigure[System model.]
    {
        \includegraphics[height=0.17\textwidth]{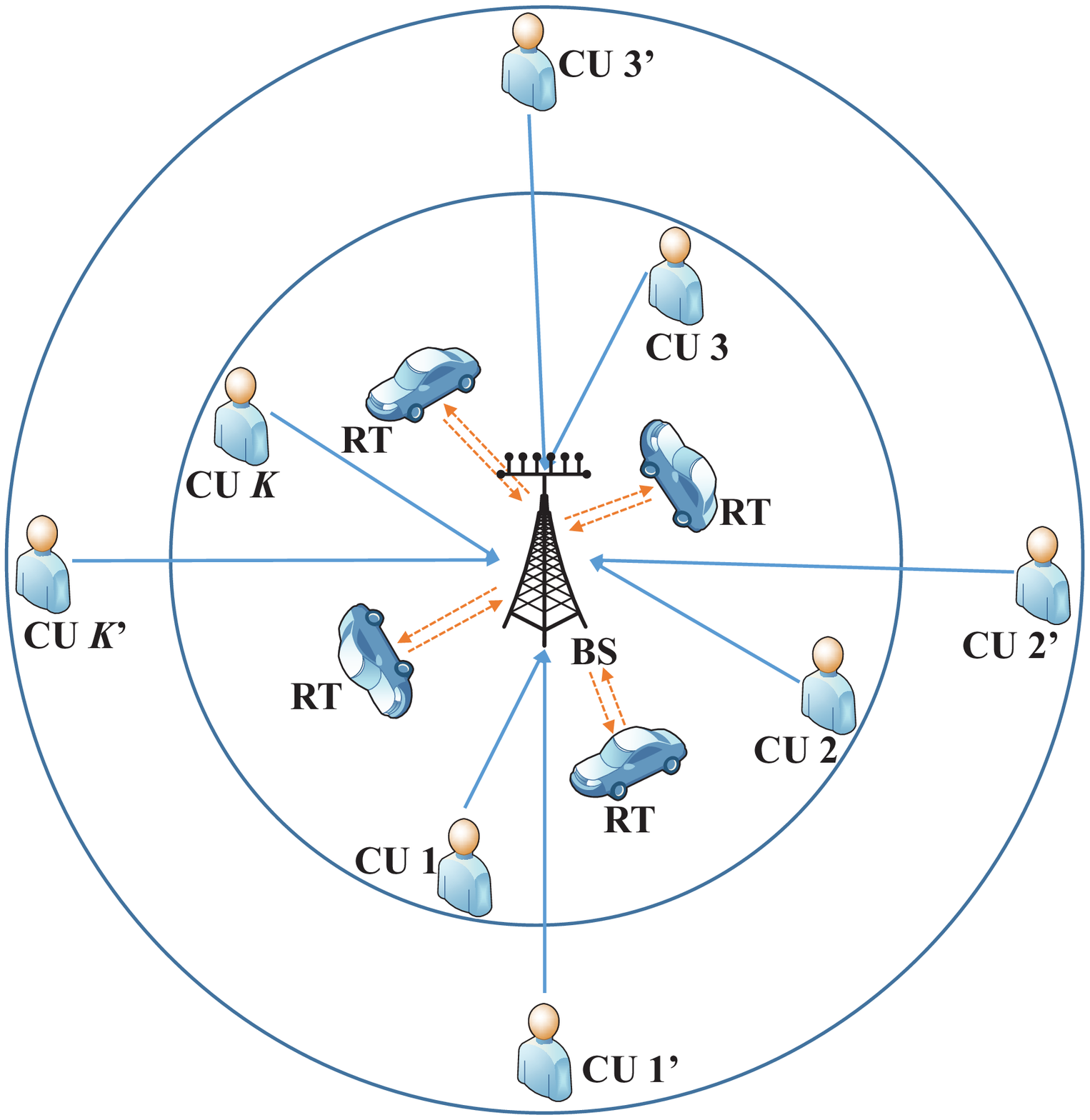}
	   \label{System_Model}	
    }
    \subfigure[RadCom BS.]
    {
        \includegraphics[height=0.17\textwidth]{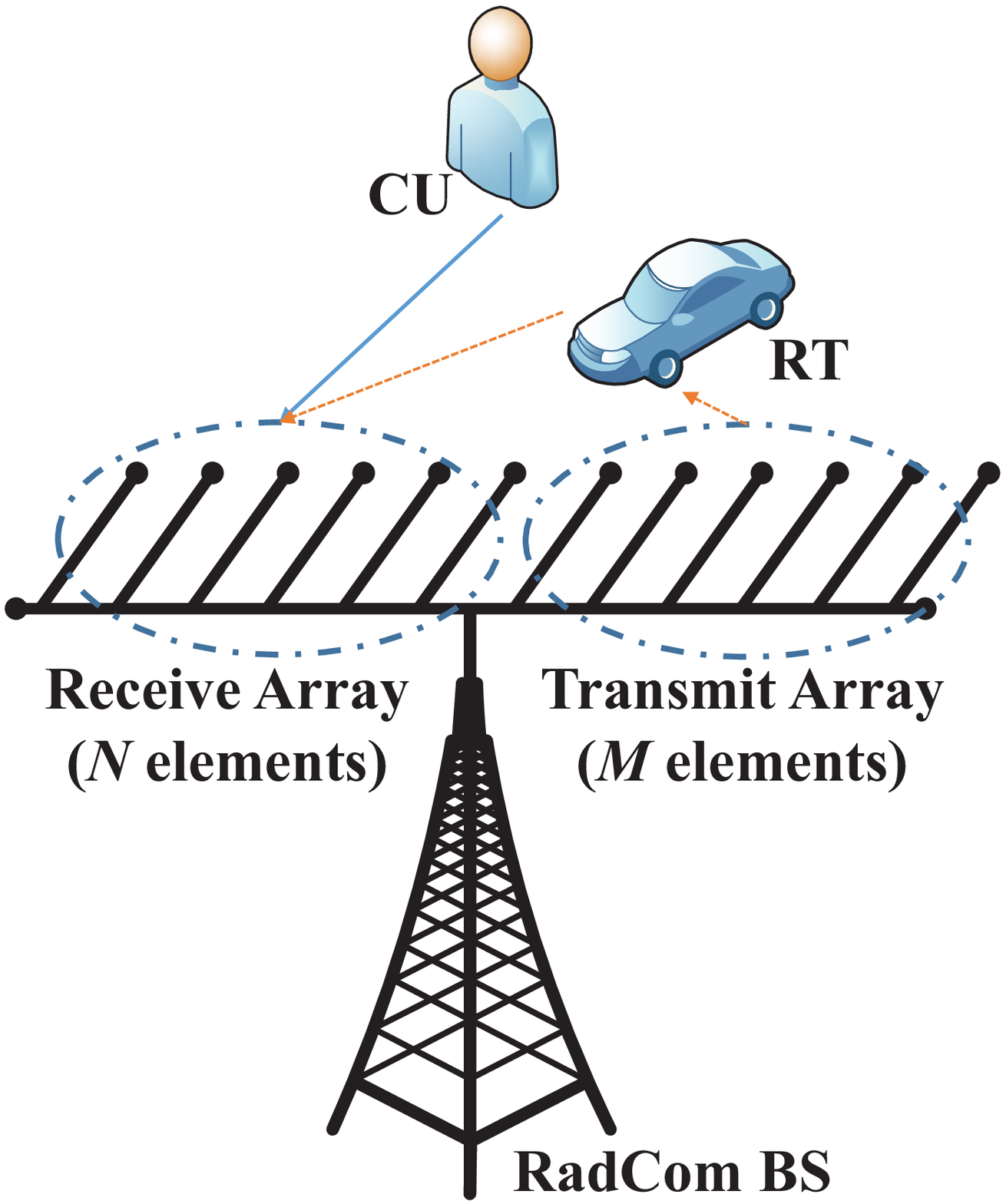}
	   \label{RadCom_BS}	
    }
\caption{An ISAC system with $2K$ CUs and several RTs.}
    \label{figure1}
    \vspace{-15pt}
\end{figure}

In an uplink ISAC system shown in {\figurename} \ref{System_Model}, one radar-communications (RadCom) base station (BS) receives data from $2K$ single-antenna communication users while simultaneously sensing the radar targets (RTs). There are two types of CUs, namely near CUs and cell-edge CUs. Particularly, the $2K$ CUs communicate with the BS under a signal alignment-based NOMA protocol \cite{Ding2016}. Moreover, the CUs are grouped into $K$ groups and the $k$th group contains one near CU $k$ and one cell-edge CU $k'$ \cite{Ding2016}. For sensing, the BS should broadcast a radar waveform to the nearby environment and extract environmental information from the radar waveform reflected by the RTs. Since the time interval of a radar waveform may be longer than the round-trip time of radar waveform travelling between the BS and the RTs, the BS should work on the full duplex mode \cite{Rahman2020}. To mitigate the resultant self-interference, we consider that the BS is equipped with two sets of spatially well-separated antennas \cite{Liu2022,Rahman2020,Tang2019}, i.e., $M$ transmit antennas and $N$ receive antennas, whose structure is illustrated in {\figurename} \ref{RadCom_BS}. For simplicity, we assume the self-interference can be completely eliminated \cite{Liu2022,Rahman2020,Tang2019}.

Let ${\textbf{S}}=\left[{\textbf{s}}_1 \cdots {\textbf{s}}_L\right]\in{\mathbbmss{C}}^{M\times L}$ ($L\geq M$, $L\geq N$) denote the radar waveform sent by the BS, where $L$ denotes the length of the radar waveform and ${\textbf{s}}_l\in{\mathbbmss C}^{M\times1}$ denotes the waveform at the $l$th time slot. Besides, the waveform is subject to the power budget ${\mathsf{tr}}\left({\textbf{S}}{\textbf{S}}^{\mathsf{H}}\right)\leq p_{\text{s}}$ with $p_{\text{s}}$ denoting the sensing SNR. In this letter, we consider the case where the BS receives the communication messages sent by CUs and the reflected radar waveform simultaneously. For brevity, we assume the BS knows the perfect channel state information of CUs and can remove the radar waveform reflected by CUs perfectly \cite{Liu2022}. Hence, the signal received by the BS is given by
\begin{align}\label{RadCom_Signal_Matrix}
{\textbf{Y}}=\sum\nolimits_{k=1}^{K}\left(\sqrt{\alpha_k}{\textbf{h}}_k{\textbf{x}}_k^{\mathsf{H}}+\sqrt{\alpha_{k'}}{\textbf{h}}_{k'}{\textbf{x}}_{k'}^{\mathsf{H}}\right)+
{\textbf{G}}^{\mathsf{H}}{\textbf{S}}+{\textbf{N}},
\end{align}
where $\alpha_i$ and ${\textbf{h}}_i\sim{\mathcal{CN}}\left({\textbf{0}},{\textbf{I}}_N\right)$ ($i\in\{k,k'\}$) model the influence of large-scale path loss and small-scale fading from CU $i$ to the BS, respectively; ${\textbf{x}}_i=\left[x_{i,1},\cdots,x_{i,L}\right]^{\mathsf{H}}$ ($i\in\{k,k'\}$) is the message sent by CU $i$ with ${\mathbbmss{E}}\left\{\left|x_{i,l}\right|^2\right\}=p_{\text{c}}$ being the transmit SNR; ${\textbf{N}}=\left[{\textbf{n}}_1 \cdots {\textbf{n}}_L\right]\in{\mathbbmss{C}}^{N\times L}$ denotes the additive white Gaussian noise (AWGN) with ${\textbf{n}}_l\sim{\mathcal{CN}}\left({\textbf{0}},{\textbf{I}}_N\right)$; $\textbf{G}=\left[g_{i,j}\right]\in{\mathbbmss{C}}^{M\times N}$ is the target response matrix to be sensed with $g_{i,j}$ representing the target response from the $i$th transmit antenna to the $j$th receive antenna \cite{Tang2019}. The target response matrix can be written as ${\textbf{G}}=\sum\nolimits_{l}\beta_{l}{\textbf{a}}\left(\theta_l\right){\textbf{b}}^{\mathsf{T}}\left(\theta_l\right)$ \cite{Tang2019}, where $\beta_{l}\sim{\mathcal{CN}}\left(0,\sigma_l^2\right)$ is the complex amplitude of the $l$th RT with $\sigma_l^2$ representing the average strength, ${\textbf{a}}\left(\theta_l\right)\in{\mathbbmss{C}}^{M\times 1}$ and ${\textbf{b}}\left(\theta_l\right)\in{\mathbbmss{C}}^{N\times 1}$ are the associated transmit and receive array steering vectors, respectively, and $\theta_l$ is its direction of arrival. Note that sensing or positioning the RTs can be regarded as estimating the response matrix $\textbf{G}$ \cite{Tang2019}. Having $\textbf{G}$ at hand, one can perform further analyses to extract more information from $\textbf{G}$, such as the direction and reflection coefficient of each RT \cite{Liu2022}. Assuming that the receive antennas of the BS are widely separated, we can simplify the response matrix to ${\textbf{G}}^{\mathsf{H}}={\textbf{Q}}{\textbf{R}}_{\text{T}}^{1/2}$ \cite{Tang2019}, where ${\textbf{R}}_{\text{T}}\in{\mathbbmss{C}}^{M\times M}$ is the transmit correlation matrix and ${\textbf{Q}}\in{\mathbbmss{C}}^{N\times M}$ contains $NM$ independent and identically distributed elements with zero mean and unit variance. Since $\textbf{G}$ needs to be sensed, we only assume the correlation matrix ${\textbf{R}}_{\textbf{T}}$ is known by the RadCom BS \cite{Tang2019}.

After the BS receives the superposed signal matrix $\textbf{Y}$, it can leverage a successive interference cancellation (SIC)-based framework to decode the communication signals, ${\textbf{x}}_i$, as well as sensing the response matrix $\textbf{G}$ \cite{Chiriyath2016}. Particularly, the BS first decodes ${\textbf{x}}_i$ by treating the radar waveform as interference. Then, ${\textbf{x}}_i$ can be subtracted from $\textbf{Y}$ and the rest part will be used for sensing $\textbf{G}$\footnote{In fact, it makes sense to consider another SIC order where $\textbf{G}$ is estimated firstly by treating ${\textbf{x}}_i$ as interference. Yet, in this case, it is challenging to quantify the influence of ${\textbf{x}}_i$ on estimating $\textbf{G}$ as well as to evaluate the statistics of the residual error in estimating $\textbf{G}$, which makes the subsequent analyses intractable. Thus, in this letter, we adopt the SIC order utilized in \cite{Chiriyath2016} and detect ${\textbf{x}}_i$ firstly. A more detailed discussion on the influence of the SIC order on the performance of the ISAC will be left to our future works.}. In the sequel, we will discuss the performance of communications and sensing, respectively.

\section{Performance of Communication Signals}
\subsection{Inter-Group Interference Cancellation}
For simplicity, we only focus on the $i$th time slot. The filtering matrix ${\textbf{W}}=\left[{\textbf{w}}_1,\cdots,{\textbf{w}}_K\right]\in{\mathbbmss C}^{N\times K}$ is adopted to decode the messages transmitted by each group in parallel. In this case, the data stream sent by the $k$th group is given by
\begin{equation}
\begin{split}
y_{k,i}&=r_{k,i}+{\textbf{w}}_k^{\mathsf{H}}\left(\sqrt{\alpha_k}{\textbf{h}}_kx_{k,i}+\sqrt{\alpha_{k'}}{\textbf{h}}_{k'}x_{{k'},i}\right)\\
&+\sum\nolimits_{j\neq k}{\textbf{w}}_k^{\mathsf{H}}\left(\sqrt{\alpha_j}{\textbf{h}}_jx_{j,i}+\sqrt{\alpha_{j'}}{\textbf{h}}_{j'}x_{{j'},i}\right)+n_{k,i},
\end{split}
\end{equation}
where $r_{k,i}\triangleq{\textbf{w}}_k^{\mathsf{H}}{\textbf{G}}^{\mathsf{H}}{\textbf{s}}_i$ and $n_{k,i}\triangleq {\textbf{w}}_k^{\mathsf{H}}{\textbf{n}}_i$. To avoid the inter-group interference (IGI), we assume $N\geq 2K$ and resort to the zero-forcing (ZF) combiner \cite{Ding2016}. In this case, we have ${\textbf{w}}_k=\frac{{\textbf{p}}_k}{\left\|{\textbf{p}}_k\right\|}$, where ${\textbf{p}}_k$ denotes the $\left(2k-1\right)$th column of matrix ${\textbf{P}}_k={\textbf{H}}_k\left({\textbf{H}}_k^{\mathsf{H}}{\textbf{H}}_k\right)^{-1}$ with ${\textbf{H}}_k$ being the sub-matrix of ${\textbf{H}}= \left[{\textbf{h}}_1 {\textbf{h}}_{1'}\cdots{\textbf{h}}_K {\textbf{h}}_{K'}\right]$ after removing the $\left(2k\right)$th column, ${\textbf{h}}_{k'}$. Since ${\textbf{I}}_{2K-1}={\textbf{H}}_k^{\mathsf{H}}{\textbf{P}}_k$, we can simplify $y_{k,i}$ as
\begin{equation}\label{CU_Signal}
\begin{split}
y_{k,i}\!=\!{\textbf{w}}_k^{\mathsf{H}}\left(\sqrt{\alpha_k}{\textbf{h}}_kx_{k,i}\!+\!\sqrt{\alpha_{k'}}{\textbf{h}}_{k'}x_{{k'},i}\right)\!+\!r_{k,i}\!+\!n_{k,i}.
\end{split}
\end{equation}
Afterwards, the BS can decode the message $\left(x_{k,i},x_{k',i}\right)$ by employing SIC \cite{Ding2016}. Notably, for uplink NOMA, the sum rate of each user group is always the same, no matter which decoding order is used \cite{Ding2016}. Thus, consider that $x_{k,i}$ is decoded at first. From a worst-case design perspective \cite{Hassibi2003}, the aggregate interference-plus-noise $\sqrt{\alpha_{k'}}{\textbf{w}}_k^{\mathsf{H}}{\textbf{h}}_{k'}x_{k',i}+r_{k,i}+n_{k,i}$ is treated as the Gaussian noise. Besides, it is worth mentioning that the RadCom BS has no full CSI of ${\textbf{G}}$. Consequently, within the $k$th group, $x_{k,i}$ can be decoded with the signal-to-interference-plus-noise ratio (SINR) $\gamma_{k,i}=\frac{\alpha_k p_{\text{c}}\left|{\textbf{w}}_k^{\mathsf{H}}{\textbf{h}}_k\right|^2}{\alpha_{k'}p_{\text{c}}\left|{\textbf{w}}_k^{\mathsf{H}}{\textbf{h}}_{k'}\right|^2
+{\mathbbmss{E}}\left\{\left|r_{k,i}\right|^2
+\left|n_{k,i}\right|^2\right\}}$, whereas $x_{k',i}$ can be decoded with the SINR $\gamma_{k',i}=\frac{\alpha_{k'}p_{\text{c}}\left|{\textbf{w}}_k^{\mathsf{H}}{\textbf{h}}_{k'}\right|^2}
{{\mathbbmss{E}}\left\{\left|r_{k,i}\right|^2+\left|n_{k,i}\right|^2\right\}}$. Taken together, the communication rate of CU $k$, CU $k'$, and the $k$th CU group can be written as ${\mathcal{R}}_{k,i}=\log_2\left(1+\gamma_{k,i}\right)$, ${\mathcal{R}}_{k',i}=\log_2\left(1+\gamma_{k',i}\right)$, and ${\mathcal{R}}_{{\text{g}},{k},i}={\mathcal{R}}_{k,i}+{\mathcal{R}}_{k',i}=\log_2\left(1+\gamma_{{\text{g}},{k},i}\right)$ with $\gamma_{{\text{g}},{k},i}=\frac{\alpha_{k}\left|{\textbf{w}}_k^{\mathsf{H}}{\textbf{h}}_{k}\right|^2+\alpha_{k'}\left|{\textbf{w}}_k^{\mathsf{H}}
{\textbf{h}}_{k'}\right|^2}{p_{\text{c}}^{-1}{\mathbbmss{E}}\left\{\left|r_{k,i}\right|^2+\left|n_{k,i}\right|^2\right\}}$, respectively. In the sequel, we intend to discuss the outage probability and the ergodic rate achieved by communication signals.

\subsection{Outage Probability}
The OPs of CU $k$, CU $k'$, and the $k$th group are given by
\begin{subequations}\label{OP_Calculation_Expression}
\begin{align}
&P_{k,i}=\Pr\left(\gamma_{k,i}<\bar\gamma_{k,i}\right),\\
&P_{k',i}=1-\Pr\left(\gamma_{k,i}>\bar\gamma_{k,i},\gamma_{k',i}>\bar\gamma_{k',i}\right),\\
&P_{{\text{g}},{k},i}=\Pr\left(\gamma_{{\text{g}},{k},i}<\bar\gamma_{{\text{g}},{k},i}\right),
\end{align}
\end{subequations}
respectively, where $\bar\gamma_{k,i}=2^{\bar{\mathcal{R}}_{k,i}}-1$, $\bar\gamma_{k',i}=2^{\bar{\mathcal{R}}_{k',i}}-1$, and $\bar\gamma_{{\text{g}},{k},i}=2^{\bar{\mathcal{R}}_{{\text{g}},{k},i}}-1$ with $\bar{\mathcal{R}}_{k,i}$, $\bar{\mathcal{R}}_{k',i}$, and $\bar{\mathcal{R}}_{{\text{g}},{k},i}$ being the target rates of CU $k$, CU $k'$, and the $k$th group, respectively. The following theorem provides closed-form expressions for these OPs as well as their high-SNR approximations.
\vspace{-5pt}
\begin{theorem}\label{Theorem_CR_OP}
The OPs $P_{k,i}$, $P_{k',i}$, and $P_{{\text{g}},{k},i}$ are given by
\begin{subequations}\label{OP_Analytical_Expression}
\begin{align}
&P_{k,i}=1-{\emph{e}}^{-\frac{{\bar\gamma}_{k,i}}{\alpha_k\rho}}\sum_{l=0}^{N_K}\sum_{n=0}^{l}
\frac{\alpha_k^{-l}\alpha_{k'}^{n}{\rho}^{n-l}{\bar\gamma}_{k,i}^{l}}{\left(l-n\right)!n!{\left(\epsilon_k{\bar\gamma}_{k,i}+1\right)^{n+1}}},\\
&P_{k',i}=1-\sum_{l=0}^{N_K}\sum_{n=0}^{l}
\frac{\frac{{\bar\gamma}_{k,i}^{l}\alpha_{k'}^{n}}{{\rho}^{l-n}}\Gamma\left(n+1,\frac{{\bar\gamma_{k',i}}\left(\epsilon_k{\bar\gamma}_{k,i}+1\right)}{\alpha_{k'}\rho}\right)}
{\alpha_k^{l}{\emph{e}}^{\frac{{\bar\gamma}_{k,i}}{\alpha_k\rho}}\left(l-n\right)!n!{\left(\epsilon_k{\bar\gamma}_{k,i}\!+\!1\right)^{n+1}}},\\
&P_{{\text{g}},{k},i}\approx\sum_{l=0}^{T}\frac{\left(1-\epsilon_k\right)^l\epsilon_k^{N_K+1}
\gamma\left(l+N_K+2,\frac{\bar\gamma_{{\text{g}},{k},i}}{\alpha_{k'}\rho}\right)}{l!N_K!\left(l+N_K+1\right)}
,\label{OP_Analytical_Expression_3}
\end{align}
\end{subequations}
respectively, where $N_K=N-2K+1$, $\rho=\frac{p_{\text{c}}}{\sigma^2}$, $\epsilon_k=\alpha_{k'}/\alpha_k$, $\sigma^2=1+\left|{\textbf{s}}_i^{\mathsf{H}}{\textbf{R}}_{\text{T}}{\textbf{s}}_i\right|$, $T$ is a complexity-vs-accuracy tradeoff parameter, $\Gamma\left(s,x\right)=\int_{x}^{\infty}t^{s-1}{\emph{e}}^{-t}{\rm{d}}t$ denotes the upper incomplete gamma function \cite[eq. 8.350.2]{Ryzhik2007}, and $\gamma\left(s,x\right)=\int_{0}^{x}t^{s-1}{\emph{e}}^{-t}{\rm{d}}t$ denotes the lower incomplete gamma function \cite[eq. 8.350.1]{Ryzhik2007}. When $p_{\text{c}}\rightarrow\infty$, the OPs satisfy:
\begin{subequations}\label{OP_Asym}
\begin{align}
&\lim_{p_{\text{c}}\rightarrow\infty}P_{k,i}=\lim_{p_{\text{c}}\rightarrow\infty}P_{k',i}=1-\sum\nolimits_{l=0}^{N_K}\frac{\epsilon_k^l{\bar\gamma}_{k,i}^l/l!}{\left(\epsilon_k{\bar\gamma}_{k,i}\!+\!1\right)^{l+1}},\label{OP_Asym_1}\\
&\lim_{p_{\text{c}}\rightarrow\infty}P_{{\text{g}},{k},i}=\left(\frac{{\bar\gamma_{{\text{g}},{k},i}}\sigma^2}{\alpha_{k'}p_{\text{c}}}\right)^{N-2K+3}\frac{\epsilon_k^{N-2K+2}}{\left(N-2K+3\right)!}.\label{OP_Asym_3}
\end{align}
\end{subequations}
\end{theorem}
\vspace{-5pt}
\begin{IEEEproof}
Please refer to Appendix \ref{Proof_Theorem_CR_OP} for more details.
\end{IEEEproof}
\vspace{-5pt}
\begin{remark}
The OPs of CU $k$ and CU $k'$ converge to the same floor in the high-SNR regime due to the principle of uplink NOMA, indicating their diversity orders are both zero.
\end{remark}
\vspace{-5pt}
\vspace{-5pt}
\begin{remark}
A diversity order of $N-2K+3$ is achievable for the sum rate in the $k$th CU group, which can be improved by increasing the number of receive antennas, $N$.
\end{remark}
\vspace{-5pt}

\subsection{Ergodic Communication Rate}
The ECRs of CU $k$, CU $k'$, and the $k$th group are given by $\bar{\mathcal{R}}_{k,i}={\mathbbmss E}\left\{{\mathcal{R}}_{k,i}\right\}$, $\bar{\mathcal{R}}_{k',i}={\mathbbmss E}\left\{{\mathcal{R}}_{k',i}\right\}$, and $\bar{\mathcal{R}}_{{\text{g}},{k},i}={\mathbbmss E}\left\{{\mathcal{R}}_{{\text{g}},{k},i}\right\}$, respectively. By definition, we have $\bar{\mathcal{R}}_{{\text{g}},{k},i}=\bar{\mathcal{R}}_{k,i}+\bar{\mathcal{R}}_{k',i}$. Theorem \ref{Theorem_CR_ER} provides closed-form expressions for the ECRs as well as their high-SNR approximations.
\vspace{-5pt}
\begin{theorem}\label{Theorem_CR_ER}
The ECRs $\bar{\mathcal{R}}_{k',i}$, $\bar{\mathcal{R}}_{{\text{g}},{k},i}$, and $\bar{\mathcal{R}}_{k,i}$ are given by
\begin{subequations}\label{ECR_Analytical_Results}
\begin{align}
&\bar{\mathcal{R}}_{k',i}=-{\emph{e}}^{1/(\beta_k\rho)}{\rm{Ei}}\left(-1/(\alpha_{k'}\rho)\right)\log_2{\emph{e}},\\
&\bar{\mathcal{R}}_{{\text{g}},{k},i}\!\approx\!f_{{\text{g}},{k},i}^{T}\!=\!\sum_{l=0}^{T}\!\sum_{n=0}^{l+\bar{N}_K}\!\!\frac{\left(1-\epsilon_k\right)^l\epsilon_k^{\bar{N}_K}\Gamma\left(l+\bar{N}_K\right)\log_2{\emph{e}}}
{l!N_K!\left(\!l\!+\!\bar{N}_K\!-\!n\!\right)!(\!-\rho\alpha_{k'}\!)^{{\bar{N}_K+l-n}}}\nonumber\\
&\times\!\!\left[\!-{\emph{e}}^{\frac{1}{\rho\alpha_{k'}}}\!{\rm{Ei}}\left(\frac{-1}{\rho\alpha_{k'}}\right)\!+\!\sum_{j=1}^{\bar{N}_K+l-n}\!\left(j-1\right)!\left(\!\frac{-1}{\rho\alpha_{k'}}\!\right)^{-j}\!\right],\\
&\bar{\mathcal{R}}_{k,i}\approx f_{{\text{g}},{k},i}^{T}-\bar{\mathcal{R}}_{k',i},
\end{align}
\end{subequations}
respectively, where $\bar{N}_K=N-2K+2$ and ${\rm{Ei}}\left(x\right)=-\int_{-x}^{\infty}{\emph{e}}^{-t}t^{-1}{\rm{d}}t$ denotes the exponential integral function \cite[eq. (8.211.1)]{Ryzhik2007}. If the transmit SNR $p_{\text{c}}$ approaches infinity, the ECRs can be approximated as follows:
\begin{subequations}\label{ECR_Asym}
\begin{align}
&\bar{\mathcal{R}}_{k,i}\approx\sum\nolimits_{l=0}^{T}\!\frac{\Gamma\left(l\!+\!\bar{N}_K\right)\psi\left(l\!+\!\bar{N}_K\!+\!1\right)}
{\left(1\!-\!\epsilon_k\right)^{-l}l!N_K!\epsilon_k^{-\bar{N}_K}\ln{2}}-\frac{\psi\left(1\right)}{\ln{2}},\label{ECR_Asym1}\\
&\bar{\mathcal{R}}_{k',i}\approx\log_2\left(\rho\alpha_{k'}\right)+\psi\left(1\right)\log_2{\emph{e}},\label{ECR_Asym2}\\
&\bar{\mathcal{R}}_{{\text{g}},{k},i}\approx\log_2\left(\rho\alpha_{k'}\right)\!+\!\sum_{l=0}^{T}\!\frac{\Gamma\left(l\!+\!\bar{N}_K\right)\psi\left(l\!+\!\bar{N}_K\!+\!1\right)}
{\left(1\!-\!\epsilon_k\right)^{-l}l!N_K!\epsilon_k^{-\bar{N}_K}\ln{2}}\label{ECR_Asym3},
\end{align}
\end{subequations}
respectively, where $\psi\left(x\right)\!=\!\frac{{\rm d}}{{\rm d}x}\ln{\Gamma\left(x\right)}$ is the Digamma function \cite[eq. (6.461)]{Ryzhik2007} and $\Gamma\left(x\right)=\int_{0}^{\infty}t^{x-1}{\emph e}^{-t}{\rm d}t$ is the gamma function \cite[eq. (6.1.1)]{Ryzhik2007}.
\end{theorem}
\vspace{-5pt}
\begin{IEEEproof}
Please refer to Appendix \ref{Proof_Theorem_CR_ER} for more details.
\end{IEEEproof}
\vspace{-5pt}
\begin{remark}
The high-SNR slopes of CU $k$, CU $k'$, and the $k$th CU group are given by 0, 1, 1, respectively, which are not affected by the number of RadCom BS antennas.
\end{remark}
\vspace{-5pt}
\vspace{-5pt}
\begin{corollary}
At the $i$th time slot, the ergodic sum communication rate of the $2K$ CUs satisfies $\bar{\mathcal{R}}_i=\sum_{k=1}^{K}\bar{\mathcal{R}}_{{\text{g}},{k},i}\approx K\left(\log_2\left(\rho\alpha_{k'}\right)\!+\!\sum_{l=0}^{T}\!\frac{\Gamma\left(l\!+\!\bar{N}_K\right)\psi\left(l\!+\!\bar{N}_K\!+\!1\right)}
{\left(1\!-\!\epsilon_k\right)^{-l}l!N_K!\epsilon_k^{-\bar{N}_K}\ln{2}}\right)$ as $p_{\text{c}}\rightarrow\infty$.
\end{corollary}
\vspace{-5pt}
\vspace{-5pt}
\begin{remark}
A high-SNR slope of $K$ is achievable for the uplink sum rate of the $K$ CU groups.
\end{remark}
\vspace{-5pt}
\vspace{-5pt}
\begin{remark}
By setting $\alpha_{k'}=0$, Theorem \ref{Theorem_CR_OP} and Theorem \ref{Theorem_CR_ER} can apply to ISAC systems where an orthogonal multiple access (OMA) protocol is involved in the communication procedure.
\end{remark}
\vspace{-5pt}
\subsection{Communication Performance of FDSAC}
In this part, we consider FDSAC as a baseline scenario, where the total bandwidth is partitioned into two sub-bands according to some $\alpha$, one for radar only and the other for communications. Particularly, we assume $\alpha$ fraction of the total bandwidth is used for communications with $\alpha\in\left[0,1\right]$. Under this circumstance, the ergodic sum rate of the $K$ groups can be written as $\bar{\mathcal{R}}_{\text{c,f}}^{\alpha}={\mathbbmss{E}}\left\{{\mathcal{R}}_{\text{c,f}}^{\alpha}\right\}$ with ${\mathcal{R}}_{\text{c,f}}^{\alpha}=\sum_{k=1}^{K}\alpha\log_2\left(1+\left(\alpha_k\left|{\textbf{w}}_k^{\mathsf{H}}{\textbf{h}}_{k}\right|^2+\alpha_{k'}\left|{\textbf{w}}_k^{\mathsf{H}}
{\textbf{h}}_{k'}\right|^2\right)p_{\text{c}}/\alpha\right)$. It is worth noting that $\bar{\mathcal{R}}_{\text{c,f}}$ can be analyzed by following a similar approach as that in analyzing $\bar{\mathcal{R}}_i$.

\section{Performance of Sensing Signals}
After decoding all the information bits sent by the CUs, the BS can remove the communication signal from the received superposed signal matrix in \eqref{RadCom_Signal_Matrix}, and then the rest part can be used for radar sensing \cite{Chiriyath2016}, which is expressed as
\begin{align}
{\textbf{Y}}_{\text{s}}={\textbf{G}}^{\mathsf{H}}{\textbf{S}}+{\textbf{N}}.
\end{align}
Since ${\textbf{Y}}_{\text{s}}$ and ${\textbf{S}}$ are both available, the RadCom BS can leverage them to sense the channel $\textbf{G}$. In this letter, we use the sensing rate to evaluate the sensing performance, which is defined as the sensing mutual information (MI) per unit time \cite{Zhang2021}. There are two reasons for using this performance metric. One is that the sensing rate tells how much information we can extract from the nearby environment from an information-theoretical point of view, the other is that maximizing the sensing rate is equivalent to minimizing the mean-square error in estimating the target response matrix $\textbf{G}$ \cite{Tang2019}. Assuming that each waveform symbol lasts 1 unit time, we can characterize the sensing rate as ${\mathcal{I}}_{L}/L$, where ${\mathcal{I}}_{L}$ denotes the sensing MI over the duration of $L$ symbols. For convenience, we consider the case of ${\textbf{R}}_{\text{T}}\succ{\textbf{0}}$ and write the sensing MI as \cite{Tang2019}
\begin{align}
{\mathcal{I}}_{L}=I\left({\textbf{Y}}_{\text{s}};{\textbf{G}}|{\textbf{S}}\right)=N\log_2\det\left({\textbf{I}}+{\textbf{S}}^{\mathsf{H}}{\textbf{R}}_{\text{T}}{\textbf{S}}\right),
\end{align}
where $I\left(X;Y|Z\right)$ denotes the MI between $X$ and $Y$ conditioned on $Z$. Thus, the maximal achievable sensing rate is
\begin{align}\label{Sensing_MI_Max}
{\mathcal{R}}_{\text{s}}=\frac{N}{L}\max_{{\mathsf{tr}}\left({\textbf{S}}{\textbf{S}}^{\mathsf{H}}\right)\leq p_{\text{s}}}\log_2\det\left({\textbf{I}}_L+{\textbf{S}}^{\mathsf{H}}{\textbf{R}}_{\text{T}}{\textbf{S}}\right).
\end{align}
Theorem \ref{Theorem_SR_ER} provides an exact expression for the maximal sensing rate as well as its high-SNR approximation.
\vspace{-5pt}
\begin{theorem}\label{Theorem_SR_ER}
The maximal achievable sensing rate of the considered ISAC system can be written as
\begin{equation}
\begin{split}
{\mathcal{R}}_{\text{s}}=\frac{N}{L}\sum\nolimits_{m=1}^{M}\log_2\left(1+\lambda_ms_m^{\star}\right),
\end{split}
\end{equation}
where $\left\{\lambda_m\right\}_{m=1}^{M}$ denote the eigenvalues of ${\textbf{R}}_{\text{T}}$ and $s_{m}^{\star}=\max\left\{0,{1}/{\nu}-{1}/{\lambda_i}\right\}$ with $\nu$ satisfying $\sum_{m=1}^{M}\max\left\{0,{1}/{\nu}-{1}/{\lambda_i}\right\}=p_{\text{s}}$. The maximal sensing rate ${\mathcal{R}}_{\text{s}}$ is achieved when the eigendecomposition (ED) of ${\textbf{S}}{\textbf{S}}^{\mathsf{H}}$ satisfies ${\textbf{S}}{\textbf{S}}^{\mathsf{H}}={\textbf{U}}_{\text{T}}^{\mathsf{H}}{\bm\Delta}^{\star}{\textbf{U}}_{\text{T}}$, where ${\textbf{U}}_{\text{T}}^{\mathsf{H}}{\mathsf{diag}}\left\{\lambda_1,\cdots,\lambda_{M}\right\}{\textbf{U}}_{\text{T}}$ denotes the ED of ${\textbf{R}}_{\text{T}}$ with $\lambda_1\geq\cdots\geq\lambda_M>0$ and ${\bm\Delta}^{\star}={\mathsf{diag}}\left\{s_1^{\star},\cdots,s_M^{\star}\right\}$.
If the sensing SNR $p_{\text{s}}$ approaches infinity, the maximal achievable sensing rate satisfies
\begin{align}
{\mathcal{R}}_{\text{s}}
\approx\frac{NM}{L}\left(\log_2{p_{\text{s}}}+\frac{1}{M}\sum\nolimits_{m=1}^{M}\log_2\left(\frac{\lambda_m}{M}\right)\right)\label{Asym_Sensing_Rate}.
\end{align}
\end{theorem}
\vspace{-5pt}
\begin{IEEEproof}
Please refer to Appendix \ref{Proof_Theorem_SR_ER} for more details.
\end{IEEEproof}
\vspace{-5pt}
\begin{remark}
A high-SNR slope of $\frac{NM}{L}$ is achievable for the maximal achievable sensing rate, which can be improved by increasing the number of BS antennas.
\end{remark}
\vspace{-5pt}
\vspace{-5pt}
\begin{corollary}\label{Corollary_Radar_Interference}
When the optimal radar waveform matrix is adopted, the interference from the radar signals to the communication signals can be simplified as follows:
\begin{equation}\label{Optimal_Radar_Waveform_Interference}
{\mathbbmss{E}}\left\{\left|{\textbf{w}}_k^{\mathsf{H}}{\textbf{G}}^{\mathsf{H}}{\textbf{s}}_i\right|^2\right\}=\left\{
\begin{array}{rcl}
s_i^{\star}\lambda_i       &      & {1\leq i\leq M}\\
0     &      & {M< i\leq L}
\end{array} \right..
\end{equation}
\end{corollary}
\vspace{-5pt}
\begin{IEEEproof}
Please refer to Appendix \ref{Proof_Corollary_Radar_Interference} for more details.
\end{IEEEproof}
\vspace{-5pt}
\begin{remark}\label{Remark_Radar_Interference}
The results in Corollary \ref{Corollary_Radar_Interference} suggest that the optimal radar waveform will not influence the quality of service of communication signals at the $i$th time slot ($L\geq i>M$).
\end{remark}
\vspace{-5pt}
In the following, we set ${\textbf{R}}_{\text{T}}={\textbf{I}}_M$ and $L=M=N$ to unveil more system insights.
\vspace{-5pt}
\begin{corollary}
When ${\textbf{R}}_{\text{T}}={\textbf{I}}_M$ and $L=M=N$, the maximal achievable sensing rate is  ${\mathcal{R}}_{\text{s}}=N\log_2\left(1+\frac{p_{\text{s}}}{N}\right)\triangleq {\mathcal{R}}(N)$.
\end{corollary}
\vspace{-5pt}
By checking the first-order derivative of ${\mathcal{R}}(x)$ with respect to $x\geq 1$, we find that $\frac{{\rm{d}}}{{\rm{d}}x}{\mathcal{R}}(x)\geq 0$ ($x\geq 1$). Additionally, it can be found that $\lim_{N\rightarrow\infty}{\mathcal{R}}(N)=p_{\text{s}}\log_2{\emph{e}}$.
\vspace{-5pt}
\begin{remark}
The fact of $\frac{{\rm{d}}}{{\rm{d}}x}{\mathcal{R}}(x)\geq 0$ ($x\geq 1$) suggests that the maximal achievable sensing rate, ${\mathcal{R}}(N)$, increases with the RadCom BS antenna number, $N$, monotonically.
\end{remark}
\vspace{-5pt}
\vspace{-5pt}
\begin{remark}
The facts of $\lim_{N\rightarrow\infty}{\mathcal{R}}(N)=p_{\text{s}}\log_2{\emph{e}}$ and $\frac{{\rm{d}}}{{\rm{d}}x}{\mathcal{R}}(x)\geq 0$ ($x\geq 1$) indicate that the achievable sensing rate is mainly limited by the power used for radar sensing.
\end{remark}
\vspace{-5pt}
Turn now to the sensing rate achieved in the FDSAC system. As mentioned earlier, in FDSCA, $(1-\alpha)$ fraction of the total bandwidth is used for radar sensing, and thus the maximal sensing rate is given by ${\mathcal{R}}_{\text{s,f}}^{\alpha}=\frac{N(1-\alpha)}{L}\max_{{\mathsf{tr}}\left({\textbf{S}}{\textbf{S}}^{\mathsf{H}}\right)\leq p_{\text{s}}}\log_2\det\left({\textbf{I}}_L+\left(1-\alpha\right)^{-1}{\textbf{S}}^{\mathsf{H}}{\textbf{R}}_{\text{T}}{\textbf{S}}\right)$. We notice that ${\mathcal{R}}_{\text{s,f}}^{\alpha}$ presents a similar form as ${\mathcal{R}}_{\text{s}}$, which can be calculated by following the steps outlined in Appendix \ref{Proof_Theorem_SR_ER}.

\section{Rate Region Characterization}
In this part, we characterize the rate region of the considered ISAC system and the baseline FDSAC system. Particularly, we use ${\mathcal{R}}^{\text{c}}$ and ${\mathcal{R}}^{\text{s}}$ to denote the achievable average ergodic sum communication rate and sensing rate of the system, respectively. Therefore, the communication-sensing rate region of the ISAC system can be characterized as
\begin{equation}\label{Rate_Regio_ISAC}
\left\{\left({\mathcal{R}}^{\text{c}},{\mathcal{R}}^{\text{s}}\right)|0\leq{\mathcal{R}}^{\text{c}}\leq{L}^{-1}\sum\nolimits_{i=1}^{L}\bar{\mathcal{R}}_i,0\leq{\mathcal{R}}^{\text{s}}\leq{\mathcal{R}}_{\text{s}}\right\},
\end{equation}
whereas the rate region of the FDSAC system satisfies
\begin{equation}\label{Rate_Regio_FDMA}
\left\{\left({\mathcal{R}}^{\text{c}},{\mathcal{R}}^{\text{s}}\right)|0\leq{\mathcal{R}}^{\text{c}}\leq{\mathcal{R}}_{\text{c,f}}^{\alpha},0\leq{\mathcal{R}}^{\text{s}}\leq{\mathcal{R}}_{\text{s,f}}^{\alpha},\alpha\in[0,1]\right\}.
\end{equation}
\section{Numerical Results}
In this section, numerical results will be used to demonstrate the performance of the ISAC system and also verify the accuracy of the developed analytical results. The complexity-vs-accuracy tradeoff parameter is $T=30$. The parameters used for simulation are listed as follows: $N=4$, $M=4$, $L=5$, $K=2$, $\alpha_k=1$ ($\forall k$), $\alpha_{k'}=0.8$ ($\forall k$), and the eigenvalues of ${\textbf{R}}_{\text{T}}$ are $\left\{5,2,1,0.5\right\}$.

\begin{figure}[!t]
    \centering
    \subfigbottomskip=0pt
	\subfigcapskip=-5pt
\setlength{\abovecaptionskip}{0pt}
    \subfigure[Outage probability.]
    {
        \includegraphics[height=0.16\textwidth]{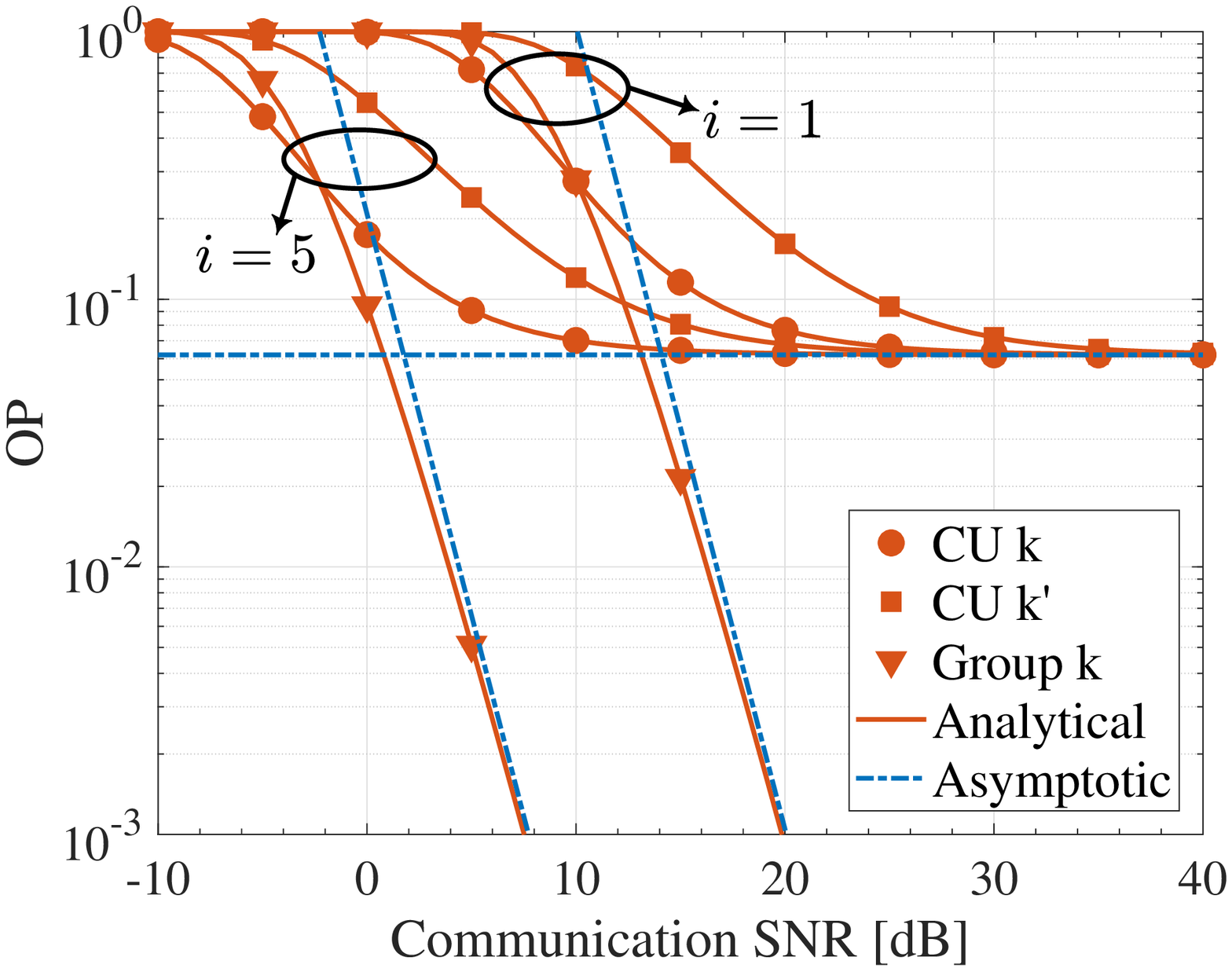}
	   \label{fig1a}	
    }
   \subfigure[Ergodic communication rate.]
    {
        \includegraphics[height=0.16\textwidth]{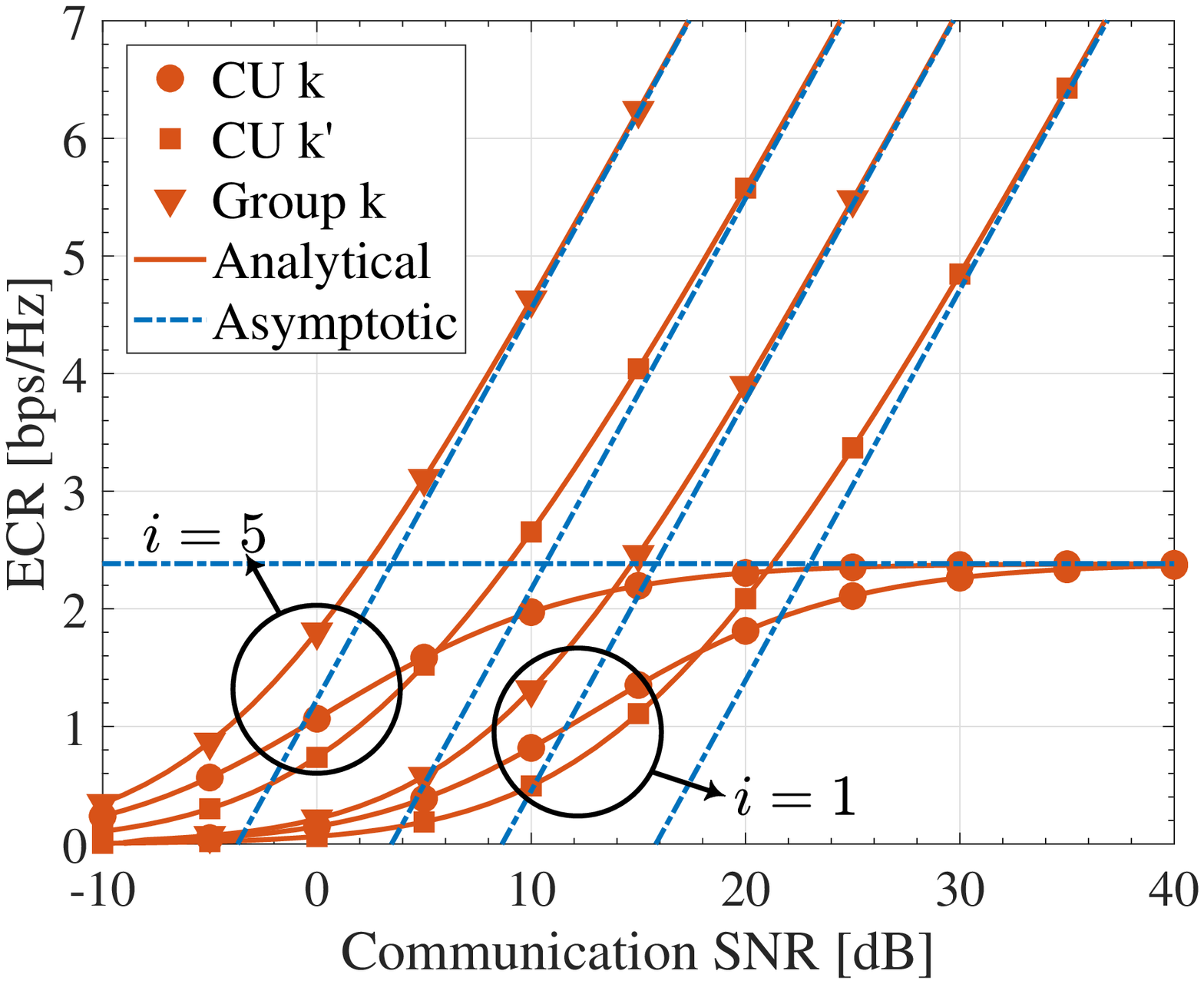}
	   \label{fig1b}	
    }
\caption{Performance of communication signals. $\bar{\mathcal{R}}_{k,i}=\bar{\mathcal{R}}_{k',i}=0.5$ bps/Hz, $\bar{\mathcal{R}}_{{\text{g}},{k},i}=1$ bps/Hz, and $p_{\text{s}}=10$ dB. The simulated results are denoted by symbols.}
    \label{figure1}
\end{figure}

\begin{figure}[!t]
 \centering
\setlength{\abovecaptionskip}{0pt}
\includegraphics[height=0.16\textwidth]{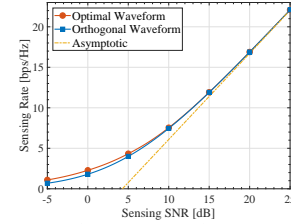}
\caption{Performance of radar signals.}
\label{figure2}
\end{figure}

{\figurename} {\ref{fig1a}} and {\figurename} {\ref{fig1b}} present the OP and ECR of CUs versus the communication SNR $p_{\text{c}}$, respectively, where the analytical results are calculated by \eqref{OP_Analytical_Expression} or \eqref{ECR_Analytical_Results} and the asymptotic results are calculated by \eqref{OP_Asym} or \eqref{ECR_Asym}. We find the analytical results fit well with the simulated results and the derived asymptotic results track the numerical results accurately in the high-SNR regime. Besides, as shown in {\figurename} {\ref{figure1}}, the ISAC system at the $5$th time slot enjoys a higher ECR as well as a lower OP than that at the $1$st time slot, which agrees with the conclusion in Remark \ref{Remark_Radar_Interference}. {\figurename} {\ref{figure2}} plots the maximal sensing rate achieved by the optimal radar waveform versus the sensing SNR $p_{\text{s}}$, where the asymptotic results are calculated with \eqref{Asym_Sensing_Rate}. We observe the asymptotic results track the numerical results accurately in the high-SNR regime. For comparison, the sensing rate achieved by the orthogonal waveform design, i.e., ${\textbf{S}}{\textbf{S}}^{\mathsf{H}}=\frac{p_{\text{s}}}{M}{\textbf{I}}_M$, is also plotted. As shown, the optimal waveform outperforms the orthogonal one in terms of the sensing rate, especially in the low-SNR region. Yet, in the high-SNR regime, these two waveforms achieve virtually the same sensing rate as well as the same high-SNR slope.

\begin{figure}[!t]
    \centering
    \subfigbottomskip=0pt
	\subfigcapskip=-5pt
\setlength{\abovecaptionskip}{0pt}
    \subfigure[$p_{\text{c}}=p_{\text{s}}=10$ dB.]
    {
        \includegraphics[height=0.16\textwidth]{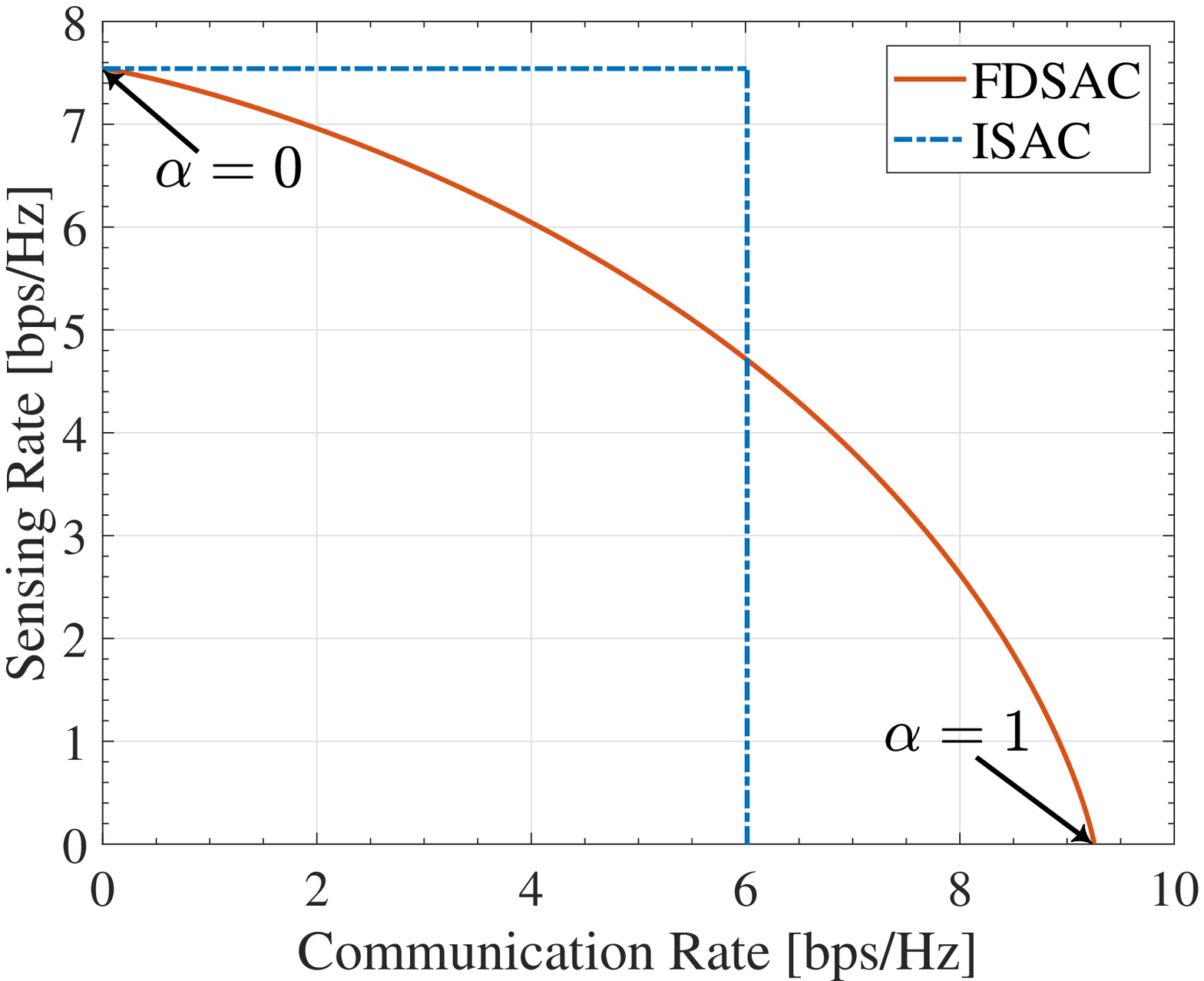}
	   \label{fig3a}	
    }
   \subfigure[$p_{\text{s}}=\beta p_{\text{c}}$ with $0\leq\beta\leq 1$.]
    {
        \includegraphics[height=0.16\textwidth]{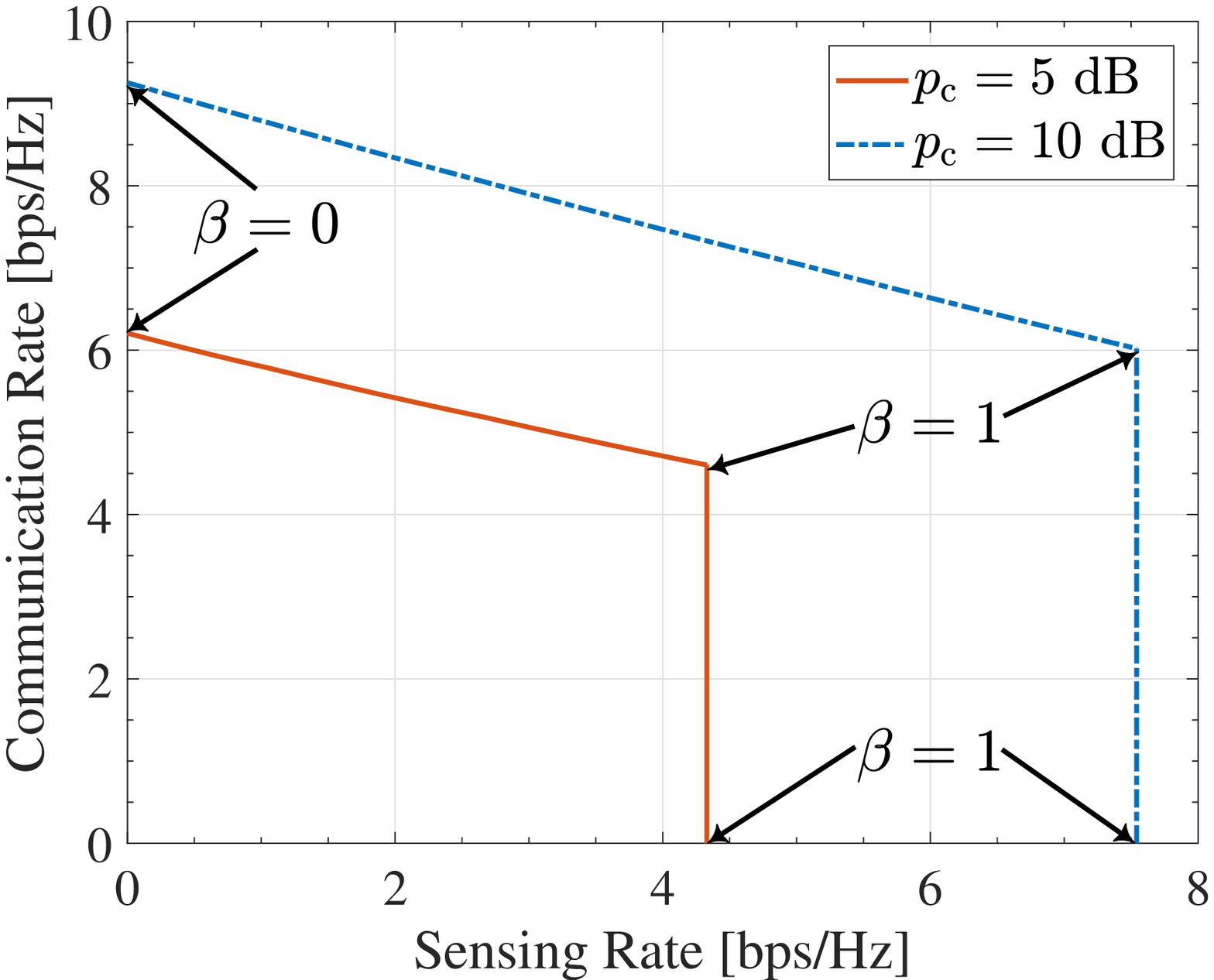}
	   \label{fig3b}	
    }
\caption{Rate region of communications and radar sensing: (a) ISAC vs FDSAC; (b) Influence of radar sensing.}
    \label{figure3}
\end{figure}

{\figurename} {\ref{fig3a}} compares the rate region of the considered ISAC system (presented in \eqref{Rate_Regio_ISAC}) and the baseline FDSAC system (presented in \eqref{Rate_Regio_FDMA}). Specifically, the rate region of the FDSAC system is plotted by changing the bandwidth allocation factor $\alpha$ from $0$ to $1$. As shown, when achieving the same communication rate, ISAC yields a higher sensing rate than FDSAC, which highlights the superiority of ISAC. In this work, we assume that the communication signals can be perfectly removed by the SIC technique before the radar sensing procedure. This means that the communication signals will not influence the sensing rate. Yet, it can be observed from \eqref{CU_Signal} that the radar signals can decrease the performance of communications. To illustrate this influence, {\figurename} {\ref{fig3b}} plots the rate region of the ISAC system when the sensing SNR satisfies $p_{\text{s}}=\beta p_{\text{c}}$ with various $0\leq\beta\leq1$. As shown, with the increment of $\beta$, the communication rate decreases monotonically, whereas the sensing rate increases, thus indicating a performance tradeoff between communications and sensing.

\section{Conclusion}
Theoretical analyses characterize the communication-sensing rate region of uplink ISAC systems. It is found that the ISAC system yields a higher sensing rate than the FDSAC system when achieving the same communication rate.

\begin{appendices}
\section{Proof of Theorem \ref{Theorem_CR_OP}}\label{Proof_Theorem_CR_OP}
\begin{IEEEproof}
We note that ${\textbf{G}}^{\mathsf{H}}{\textbf{s}}_i\!\sim\!{\mathcal{CN}}\left({\textbf{0}},\left|{\textbf{s}}_i^{\mathsf{H}}{\textbf{R}}_{\text{T}}{\textbf{s}}_i\right|
{\textbf{I}}_N\right)$, which yields ${\mathbbmss{E}}\left\{\left|r_{k,i}\right|^2+\left|n_{k,i}\right|^2\right\}=
\left|{\textbf{s}}_i^{\mathsf{H}}{\textbf{R}}_{\text{T}}{\textbf{s}}_i\right|{\textbf{w}}_k^{\mathsf{H}}{\textbf{w}}_k+1
=\left|{\textbf{s}}_i^{\mathsf{H}}{\textbf{R}}_{\text{T}}{\textbf{s}}_i\right|+1=\sigma^2$. Furthermore, the random variable $\left|{\textbf{w}}_k^{\mathsf{H}}{\textbf{h}}_k\right|^2$ follows the chi-square distribution with the PDF $f_k\left(x\right)=\frac{1}{\left(N_K\right)!}x^{N_K}{\emph{e}}^{-x}$ \cite{Lozano2018}. It is worth noting that ${\textbf{w}}_k$ is independent of ${\textbf{h}}_{k'}$, which together with the facts of $\left\|{\textbf{w}}_k\right\|=1$ and ${\textbf{h}}_{k'}\sim{\mathcal{CN}}\left({\textbf{0}},{\textbf{I}}_N\right)$, suggests that $\left|{\textbf{w}}_k^{\mathsf{H}}{\textbf{h}}_{k'}\right|^2$ is independent of $\left|{\textbf{w}}_k^{\mathsf{H}}{\textbf{h}}_k\right|^2$ and the PDF of $\left|{\textbf{w}}_k^{\mathsf{H}}{\textbf{h}}_{k'}\right|^2$ is $f_{k'}\left(x\right)={\emph{e}}^{-x}$. By \cite{Moschopoulos1985}, we find that the PDF of $\alpha_k\left|{\textbf{w}}_k^{\mathsf{H}}{\textbf{h}}_k\right|^2+\alpha_{k'}\left|{\textbf{w}}_k^{\mathsf{H}}{\textbf{h}}_{k'}\right|^2$ satisfies $f_{\bar k}\left(x\right)\approx\sum_{l=0}^{T}\frac{(1-\epsilon_k)^ly^{l+{\bar{N}_K}}{\emph{e}}^{-x/\alpha_{k'}}}{N_K!l!\alpha_k^{\bar{N}_K}\alpha_{k'}^{l+1}(l+\bar{N}_K)}$. By \eqref{OP_Calculation_Expression}, the OPs can be calculated as $P_{k,i}\!=\!\int_{0}^{\infty}\!f_{k'}\left(y\right)\!\int_{0}^{\bar\gamma_{k,i}\left(\epsilon_ky+1/(\rho\alpha_k)\right)}\!f_k\left(x\right)\!{\rm{d}}x{\rm{d}}y$, $P_{k',i}\!=\!1-\int_{\bar\gamma_{k',i}/(\rho\alpha_{k'})}^{\infty}\!f_{k'}\left(y\right)\!\int_{\bar\gamma_{k,i}\left(\epsilon_ky+1/(\rho\alpha_k)\right)}^{\infty}\!f_k\left(x\right)\!{\rm{d}}x{\rm{d}}y$, and $P_{{\text{g}},{k},i}\!=\!\int_{0}^{\bar\gamma_{{\text{g}},{k},i}/\rho}\!f_{\bar k}\left(x\right)\!{\rm{d}}x$, respectively. Inserting the obtained PDFs into the expressions of OPs and calculating the resultant integrals, we arrive at the results in \eqref{OP_Analytical_Expression}. When $p_{\text{c}}\rightarrow\infty$, using the properties of $\lim_{x\rightarrow0}{\emph{e}}^{x}=1$, $\lim_{x\rightarrow0}\Gamma\left(l+1,x\right)=l!$, and $\lim_{x\rightarrow0}\frac{\gamma(s,x)}{x^s}\rightarrow\frac{1}{s}$ \cite[eq. (8.354.1)]{Ryzhik2007}, we can get the results in \eqref{OP_Asym}.
\end{IEEEproof}
\section{Proof of Theorem \ref{Theorem_CR_ER}}\label{Proof_Theorem_CR_ER}
\begin{IEEEproof}
Inserting the obtained PDF expressions into the expressions of the ECRs and calculating the resultant integrals with the aid of \cite[eq. (4.337.5)]{Ryzhik2007}, we can obtain the results in \eqref{ECR_Analytical_Results}. By continuously using the fact of $\lim_{x\rightarrow \infty}\ln(1+x)\approx\ln{x}$ and the integral identity \cite[eq. (4.352.1)]{Ryzhik2007}, we can get the approximated results in \eqref{ECR_Asym}.
\end{IEEEproof}
\section{Proof of Theorem \ref{Theorem_SR_ER}}\label{Proof_Theorem_SR_ER}
\begin{IEEEproof}
We note that ${\mathcal{I}}_{\text{T}}\triangleq\log_2\det\left({\textbf{I}}_L+{\textbf{S}}^{\mathsf{H}}{\textbf{R}}_{\text{T}}{\textbf{S}}\right)$ can be treated as the transmission rate of a virtual MIMO channel $\bar{\textbf{y}}={\textbf{R}}_{\text{T}}^{1/2}{\textbf{S}}\bar{\textbf{x}}+\bar{\textbf{n}}$ with ${\mathbbmss{E}}\left\{\bar{\textbf{x}}{\bar{\textbf{x}}}^{\mathsf{H}}\right\}={\textbf{I}}_L$ and $\bar{\textbf{n}}\sim{\mathcal{CN}}\left({\textbf{0}},{\textbf{I}}_M\right)$. Consequently, when ${\mathcal{I}}_{\text{T}}$ is maximized, the eigenvectors of ${\textbf{S}}{\textbf{S}}^{\mathsf{H}}$ should equal the left eigenvectors of ${\textbf{R}}_{\text{T}}^{1/2}$, with the eigenvalues chosen by the water-filling procedure \cite{Lozano2018}. Hence, the maximal achievable sensing rate is given as ${\mathcal{R}}_{\text{s}}=\frac{N}{L}\sum_{m=1}^{M}\log_2\left(1+\lambda_ms_m^{\star}\right)$, where $\left\{\lambda_i\right\}_{i=1}^{M}$ denote eigenvalues of ${\textbf{R}}_{\text{T}}$ and $s_{m}^{\star}=\max\left\{0,{1}/{\nu}-{\sigma^2}/{\lambda_i}\right\}$ with $\sum_{m=1}^{M}\max\left\{0,{1}/{\nu}-{\sigma^2}/{\lambda_i}\right\}=p_{\text{s}}$. When the power used for sensing approaches infinity, namely $p_{\text{s}}\rightarrow\infty$, we have $\nu\rightarrow0$, which yields $\sum_{m=1}^{M}\max\left\{0,\frac{1}{\nu}-\frac{1}{\lambda_i}\right\}=\frac{M}{\nu}-\sum_{m=1}^{M}\frac{1}{\lambda_i}=p_{\text{s}}$. Therefore, the high-SNR sensing rate can be written as ${\mathcal{R}}_{\text{s}}=\frac{N}{L}\sum_{m=1}^{M}\log_2\left(\lambda_m/M\right)+\frac{NM}{L}\log_2\left(p_{\text{s}}+\sum_{m=1}^{M}\frac{1}{\lambda_m}\right)$. The final result follows immediately.
\end{IEEEproof}
\section{Proof of Corollary \ref{Corollary_Radar_Interference}}\label{Proof_Corollary_Radar_Interference}
\begin{IEEEproof}
By Appendix \ref{Proof_Theorem_CR_OP}, we can get ${\mathbbmss{E}}\left\{\left|r_{k,i}\right|^2\right\}
=\left|{\textbf{s}}_i^{\mathsf{H}}{\textbf{R}}_{\text{T}}{\textbf{s}}_i\right|$. Besides, the optimal waveform matrix satisfies ${\textbf{S}}{\textbf{S}}^{\mathsf{H}}={\textbf{U}}_{\text{T}}^{\mathsf{H}}{\bm\Delta}^{\star}{\textbf{U}}_{\text{T}}$, where ${\textbf{U}}_{\text{T}}^{\mathsf{H}}{\bm\Lambda}_{\text{T}}{\textbf{U}}_{\text{T}}$ denotes the ED of ${\textbf{R}}_{\text{T}}$ with ${\textbf{U}}_{\text{T}}^{\mathsf{H}}{\textbf{U}}_{\text{T}}={\textbf{I}}_{M}$ and ${\bm\Lambda}_{\text{T}}={\mathsf{diag}}\left\{\lambda_1,\cdots,\lambda_{M}\right\}$. Hence, we can get ${\textbf{S}}^{\mathsf{H}}{\textbf{R}}_{\text{T}}{\textbf{S}}={\mathsf{diag}}\left\{\lambda_1s_1^{\star},\cdots,\lambda_{M}s_M^{\star},0,\cdots,0\right\}
\in{\mathbbmss C}^{L\times L}$. The final result follows immediately.
\end{IEEEproof}
\end{appendices}


\begin{thebibliography}{00}
\bibitem{Liu2021} F. Liu, \emph{et al.}, ``Integrated sensing and communications: Towards dual-functional wireless networks for 6G and beyond,'' \emph{IEEE J. Sel. Areas Commun.}, vol. 40, no. 6, pp. 1728--1767, Jun. 2022.
\bibitem{Zhang2021} J. A. Zhang, \emph{et al.}, ``An overview of signal processing techniques for joint communication and radar sensing,'' \emph{IEEE J. Sel. Topics Signal Process.}, vol. 15, no. 6, pp. 1295--1315, Nov. 2021.
\bibitem{Mu2021} X. Mu, \emph{et al.}, ``NOMA-aided joint radar and multicast-unicast communication systems,'' \emph{IEEE J. Sel. Areas Commun.}, vol. 40, no. 6, pp. 1978--1992, Jun. 2022.
\bibitem{Wang2021} Z. Wang, \emph{et al.}, ``NOMA empowered integrated sensing and communication,'' \emph{IEEE Commun. Lett.}, vol. 26, no. 3, pp. 677--681, Mar. 2022.
\bibitem{Liu2022} F. Liu, \emph{et al.}, ``Cram\'{e}r-rao bound optimization for joint radar-communication beamforming,'' \emph{IEEE Trans. Signal Process.}, vol. 70, pp. 240--253, 2022.
\bibitem{Rahman2020} M. L. Rahman, \emph{et al.}, ``Framework for a perceptive mobile network using joint communication and radar sensing,'' \emph{IEEE Trans. Aerosp. Electron. Syst.}, vol. 56, no. 3, pp. 1926--1941, Jun. 2020.
\bibitem{Tang2019}  B. Tang and J. Li, ``Spectrally constrained MIMO radar waveform design based on mutual information,'' \emph{IEEE Trans. Signal Process.}, vol. 67, no. 3, pp. 821--834, Feb. 2019.
\bibitem{Heath2020} P. Kumari, \emph{et al.}, ``Adaptive virtual waveform design for millimeter-wave joint communication-radar,'' \emph{IEEE Trans. Signal Process}., vol. 68, pp. 715--730, 2020.
\bibitem{Chiriyath2016} A. R. Chiriyath, B. Paul, G. M. Jacyna, and D. W. Bliss, ``Inner bounds on performance of radar and communications co-existence,'' \emph{IEEE Trans. Signal Process.}, vol. 64, no. 2, pp. 464--474, Jan. 2016.
\bibitem{Ding2016} Z. Ding, \emph{et al.}, ``A general MIMO framework for NOMA downlink and uplink transmission based on signal alignment,'' \emph{IEEE Trans. Wireless Commun.}, vol. 15, no. 6, pp. 4438--4454, Jun. 2016.
\bibitem{Hassibi2003} B. Hassibi and B. M. Hochwald, ``How much training is needed in multiple-antenna wireless links?'' \emph{IEEE Trans. Inf. Theory}, vol. 49, no. 4, pp. 951--963, Apr. 2003.
\bibitem{Ryzhik2007} I. S. Gradshteyn and I. M. Ryzhik, {\emph{Table of Integrals, Series and Products}}, 7th ed., Academic, San Diego, C.A., 2007.
\bibitem{Lozano2018} R. W. Heath, Jr., and A. Lozano, \emph{Foundations of MIMO Communication}, Cambridge, U.K.: Cambridge Univ. Press, 2018.
\bibitem{Moschopoulos1985} P. G. Moschopoulos, ``The distribution of the sum of independent gamma random variables,'' \emph{Ann. Inst. Statist. Math.}, vol. 37, pp. 541--544, 1985.
\end{thebibliography}
\end{document}